\documentclass[epj,nopacs]{svjour}
\usepackage{epsfig,amsmath}

\allowdisplaybreaks[1]

\newcommand{\nn}{\nonumber}
\newcommand{\eps}{\epsilon}

\begin{document}

\title{HELAS and MadGraph/MadEvent with spin-2 particles}
\author{K.~Hagiwara\inst{1}
   \and J.~Kanzaki\inst{2}\fnmsep\thanks{e-mail: junichi.kanzaki@kek.jp}
   \and Q.~Li\inst{3}\fnmsep\thanks{e-mail:
                                       qliphy@particle.uni-karlsruhe.de}
   \and K.~Mawatari\inst{4}\fnmsep\thanks{e-mail: kentarou@kias.re.kr}
       }
\institute{KEK, Theory Division and Sokendai, Tsukuba 305-0801, Japan
      \and KEK, Tsukuba 305-0801, Japan
      \and Institut f\"ur Theoretische Physik, Universit\"at
           Karlsruhe, Postfach 6980, D-76128 Karlsruhe, Germany
      \and School of Physics, Korea Institute for Advanced Study,
           Seoul 130-722, Korea
       }
\abstract{Fortran subroutines to calculate helicity amplitudes with
massive spin-2 particles (massive gravitons), which couple to the
standard model particles via the energy momentum tensor, are added
to the {\tt HELAS} ({\tt HEL}icity {\tt A}mplitude {\tt S}ubroutines)
library. They are coded in such a way that arbitrary scattering
amplitudes with one graviton production and its decays can be generated
automatically by {\tt MadGraph} and {\tt MadEvent}, after slight
modifications. All the codes have been tested carefully by making use of
the invariance of the helicity amplitudes under the gauge and general
coordinate transformations.}

\titlerunning{HELAS and MadGraph/MadEvent with spin-2 particles}
\authorrunning{K.~Hagiwara, J.~Kanzaki, Q.~Li and K.~Mawatari}

\maketitle


\vspace*{-105mm}
\noindent KEK-TH-1218\\
\noindent KA-TP-02-2008\\
\noindent SFB/CPP-08-02\\
\noindent KIAS-P08036\\[1mm]
\today
\vspace*{83mm}

\section{Introduction}
\label{sec:1}

The idea of extra space dimensions has attracted much attention in
recent years, since it can give us a novel solution to the hierarchy
problem, or an alternative explanation of the hierarchical difference
between the Planck scale ($M_{\rm Pl}\sim 10^{19}$\,GeV) and the
electroweak scale ($m_{W}\sim 10^{2}$\,GeV).

So far, there have been various extra dimension models, which can be
divided into two major classes according to the geometry of the
background space-time manifold. The first one includes the ADD
(Arkani-Hamed, Dimopoulos, and Dvali) model~\cite{ADD} and its variants,
which extend the dimension of the total space-time to $D=4+\delta$, with
a factorizable metric and large size of the compact extra dimensions
($\gg 1/M_{\rm Pl}$). The second one includes the 5-dimensional RS
(Randall and Sundrum) model~\cite{RS} and its variants, in which a
warped metric is introduced along the 5-th dimension and the size of
the extra dimension needs not to be much larger than the Planck length.

In both classes of extra dimension models, there appear Kaluza-Klein
(KK) towers of massive spin-2 gravitons, which can interact with the
standard model (SM) fields. The effective interaction Lagrangian is
given by~\cite{Feynr1,Feynr2}
\begin{align}
 \label{IL}
 {\cal L}_{\rm int} = - \frac{1}{\Lambda}
  \sum_{\vec{n}} T^{(\vec{n})\mu \nu} {\cal T}_{\mu \nu},
\end{align}
where $T^{(\vec{n})\mu \nu}$ is the $\vec{n}$-th graviton KK modes, and
$\Lambda$ is the relevant coupling scale. In the ADD model we have
\begin{align}
 \Lambda = {\overline M}_{\rm Pl}\equiv M_{\rm Pl}/\sqrt{8\pi}
           \sim 2.4\times 10^{18}\ {\rm GeV},
\end{align}
where ${\overline M}_{\rm Pl}$ is the 4-dimensional reduced Planck
scale, and in the RS model
\begin{align}
 \Lambda=e^{-kr_c\pi}\overline{M}_{\rm Pl}
\end{align}
is at the electroweak scale, where $k$ is a scale of order of the Planck
scale and $r_c$ is the compactification radius.

In Eq.~(\ref{IL}), ${\cal T}_{\mu \nu}$ is the energy-momentum tensor of
the SM fields,
\begin{align}
\label{tensor}
 {\cal T}_{\mu\nu}
 &= \Big(-\eta_{\mu\nu} {\cal L}_{\rm SM}
   +2\frac{\delta{\cal L}_{\rm SM}}{\delta g^{\mu\nu}}\Big)
    {\Big |}_{g^{\mu\nu}=\eta^{\mu\nu}} \nn\\
 &= {\cal T}^H_{\mu\nu}+{\cal T}^F_{\mu\nu}+{\cal T}^{g}_{\mu\nu}
   +{\cal T}^{\gamma}_{\mu\nu}+{\cal T}^Z_{\mu\nu}+{\cal T}^W_{\mu\nu}
   +\cdots,
\end{align}
where $g^{\mu\nu}$ is the metric and $\eta^{\mu\nu}={\rm
diag}(1,-1,-1,-1)$ is the Minkowski value, and each energy-momentum
tensor is:
\begin{subequations}
\begin{align}
 {\cal T}^H_{\mu\nu}
  =&-\eta_{\mu\nu}
      \Big[ \frac{1}{2}\partial^\rho H\partial_\rho H
            -\frac{m^2_H}{2}H^2
            -\frac{ g^{}_{W}m^2_H}{4m_W}H^3 \nn\\
  &\qquad\quad -\frac{g^{}_{W}m^2_H}{32m^2_W}H^4
            -\frac{g^{}_{W}m_F}{2m_W}\overline{\psi}\psi H\nn\\
  &\qquad\quad +\frac{g^{}_{Z}m^{}_{Z}}{2}Z_{\mu}Z^{\mu}H
            +\frac{g^2_{Z}}{8}Z_{\mu}Z^{\mu}H^2 \nn\\
  &\qquad\quad +g^{}_{W}m^{}_{W} W^+_{\mu}W^{-\mu}H
            +\frac{g^2_{W}}{4}W^+_\mu W^{-\mu}H^2
      \Big]\nn\\
  &+\partial_\mu H\partial_\nu H
       +g^{}_{Z}m^{}_{Z}Z_\mu Z_{\nu}H
       +\frac{g^2_{Z}}{4}Z_\mu Z_{\nu}H^2 \nn\\
  &\!\!+\Big[ g^{}_{W}m^{}_{W} W^+_\mu W^{-}_\nu H
               +\frac{g^2_{W}}{4}W^+_\mu W^{-}_\nu H^2
               +(\mu\leftrightarrow\nu)
          \Big], \\
 {\cal T}^F_{\mu\nu}
  =&-\eta_{\mu\nu}
     \Big[ \overline{\psi} (i\gamma^\rho D_\rho-m_F)\psi
          -\frac{1}{2}\partial^\rho(\overline{\psi}i\gamma_\rho\psi)\nn\\
  &\qquad\quad -\Big(\frac{ g^{}_{W}}{\sqrt{2}}V_{ij}
               \overline{\psi}_{u_i}\gamma^\rho P_L\psi_{d_j}W^+_\rho\nn\\
  &\qquad\quad\quad\ +\frac{g^{}_{W}}{\sqrt{2}}U_{ij}
               \overline{\psi}_{l_i}\gamma^\rho P_L\psi_{\nu_j}W^-_\rho
    +h.c.\Big)\Big]\nn\\
  &+\Big[ \frac{1}{2}\overline{\psi}i\gamma_\mu D_\nu\psi
               -\frac{1}{4}\partial_\mu(\overline{\psi}i\gamma_\nu\psi)
               +(\mu\leftrightarrow\nu)
          \Big] \nn\\
  &+\Big[-\frac{ g^{}_{W}}{\sqrt{2}}V_{ij}
               \overline{\psi}_{u_i}\gamma_\mu P_L\psi_{d_j}W^+_\nu \nn\\
  &\quad\ \ -\frac{ g^{}_{W}}{\sqrt{2}}U_{ij}
               \overline{\psi}_{l_i}\gamma_\mu P_L\psi_{\nu_j}W^-_\nu
          +h.c. +(\mu\leftrightarrow\nu)\Big],
\label{T^F}\\
 {\cal T}^{g}_{\mu\nu}
  =&-\eta_{\mu\nu}
     \Big[-\frac{1}{4}F^{a,\rho\sigma}F^a_{\rho\sigma}
          +\partial^\rho\partial^\sigma A^a_\sigma A^a_\rho
          -\frac{1}{2}(\partial^\rho A^a_\rho)^2
     \Big] \nn\\
  &-F^{a,\rho}_\mu F^a_{\nu\rho}
         +\partial_\mu\partial^\rho A^a_\rho A^a_\nu
         +\partial_\nu \partial^\rho A^a_\rho A^a_\mu,\\
 {\cal T}^{\gamma}_{\mu\nu}
  =&-\eta_{\mu\nu}
      \Big[-\frac{1}{4}F^{\rho\sigma}F_{\rho\sigma}
           +\partial^\rho\partial^\sigma A_\sigma A_\rho
           -\frac{1}{2}(\partial^\rho A_\rho)^2\Big] \nn\\
  &-F^{\ \rho}_\mu F_{\nu\rho}
         +\partial_\mu\partial^\rho A_\rho A_\nu
         +\partial_\nu \partial^\rho A_\rho A_\mu,\\
 {\cal T}^{Z}_{\mu\nu}
  =&-\eta_{\mu\nu}
      \Big[-\frac{1}{4}Z^{\rho\sigma}Z_{\rho\sigma}
           +\frac{m^2_Z}{2}Z^\rho Z_\rho\Big] \nn\\
  &-Z^{\ \rho}_\mu Z_{\nu\rho} +m^2_ZZ_\mu Z_\nu, \\
 {\cal T}^{W}_{\mu\nu}
  =&-\eta_{\mu\nu}
      \Big[-\frac{1}{2}W^{+\rho\sigma}W^-_{\rho\sigma}
           +m^2_W W^{+\rho} W_\rho^-\Big] \nn\\
  &-\big[ W^{+\rho}_\mu W^-_{\nu\rho}-m^2_WW^+_\mu W^-_\nu
         +(\mu\leftrightarrow\nu)\big],
\end{align}
\end{subequations}
with
$e=g^{}_{W}\sin\theta^{}_{W}=g^{}_{Z}\sin\theta^{}_{W}\cos\theta^{}_{W}$
and the projection operator $P_L=\frac{1}{2}(1-\gamma_5)$.
Here, the covariant derivative is
\begin{align}
 D_\mu \equiv\partial_\mu &+ig_sT^aA^a_\mu+ieQ_fA_\mu \nn\\
       &+ig^{}_{Z}(T^3P_L-Q_f\sin^2\theta^{}_{W})Z_\mu.
\end{align}
Note that the derivative couplings of the $W$ bosons are written
explicitly in Eq.~(\ref{T^F}). Each field-strength tensor for the gauge
bosons is
\begin{align}
 F^a_{\mu\nu}
  &= \partial_\mu A_\nu^a-\partial_\nu A_\mu^a
    -g_sf^{abc}A_\mu^b A_\nu^c, \nn\\
 F_{\mu\nu}
  &= \partial_\mu A_\nu-\partial_\nu A_\mu
    +ie\big[W^+_\mu W^-_\nu-(\mu\leftrightarrow\nu)\big],\nn\\
 Z_{\mu\nu}
  &= \partial_\mu Z_\nu-\partial_\nu Z_\mu
    +ig^{}_{W}\cos\theta^{}_{W}
       \big[W^+_\mu W^-_\nu-(\mu\leftrightarrow\nu)\big], \nn\\
 W^{\pm}_{\mu\nu}
  &= \partial_\mu W^{\pm}_\nu-\partial_\nu W^{\pm}_\mu \nn\\
  &\quad \mp i g^{}_{W}\big[\sin\theta^{}_{W} W^{\pm}_\mu A_\nu
         +\cos\theta^{}_{W} W^{\pm}_\mu
         Z_\nu-(\mu\leftrightarrow\nu)\big].
\end{align}
Notice that as in the standard {\tt HELAS} package~\cite{helas}, we
use the unitary gauge for the massive vector-boson propagators and
the Feynman gauge for the massless ones.

In this paper, we present new {\tt HELAS} subroutines~\cite{helas} for
the massive gravitons and their interactions based on the effective
Lagrangian of Eq.~(\ref{IL}), and implement them into
{\tt MadGraph/MadEvent} ({\tt MG/ME})~\cite{madgraph,madevent,newmad}.%
\footnote{The Fortran code for simulations of the massive gravitons is
available on the web~\cite{KEK_HMM}.}

The paper is organized as follows:
In Sect.~\ref{sec:helas} we give the new {\tt HELAS} subroutines.
In Sect.~\ref{sec:mgme} we describe how to implement amplitudes with a
massive spin-2 graviton into {\tt MG/ME}.
In Sect.~\ref{sec:sample} we give sample numerical results.
Sect.~\ref{sec:summary} contains a brief conclusion.

\section{HELAS subroutines for spin-2 particles}
\label{sec:helas}

In this section, we list the contents of all the new {\tt HELAS}
subroutines that are needed to evaluate massive spin-2 gravtion
production at hadron colliders in association with quark and gluon jets,
and its decays into a pair of all the SM particles, or into arbitrary
numbers of quarks and gluons.

To begin with, in Sect.~\ref{sec:wavefunc} the subroutine to compute the
external lines for a spin-2 tensor particle is presented.
Next, in the following subsections, \ref{sec:vertex_i} to
\ref{sec:vertex_f}, the subroutines to compute the interactions of the
tensor boson with the SM particles are explained. The new vertex
subroutines are listed in Table~\ref{sublist}.
Those subroutines which we do not present in this paper are {\tt FFST},
{\tt SSST} and {\tt SSSST} type vertices for the interactions with
scalars; {\tt VVVT} and {\tt VVVVT} for the interactions with the
electroweak gauge bosons.
These contribute {\it e.g.} to the graviton decays into three
or more weakly interacting particles. We present the effective
Lagrangian of Eqs.~(5a) to~(5f) for completeness sake.
Finally, we briefly mention how we test our new subroutines in
Sect.~\ref{sec:check}.

\begin{table}
\centering
\begin{tabular}{|c|c|c|c|} \hline
 Vertex & Inputs & Output & Subroutine \\ \hline\hline
 SST & SST & Amplitude & {\tt SSTXXX} \\
     & ST  & S         & {\tt HSTXXX} \\
     & SS  & T         & {\tt USSXXX} \\ \hline
 FFT & FFT & Amplitude & {\tt IOTXXX} \\
     & FT  & F         & {\tt FTIXXX}, {\tt FTOXXX} \\
     & FF  & T         & {\tt UIOXXX} \\ \hline
 VVT & VVT & Amplitude & {\tt VVTXXX} \\
     & VT  & V         & {\tt JVTXXX} \\
     & VV  & T         & {\tt UVVXXX} \\ \hline
 FFVT & FFVT & Amplitude & {\tt IOVTXX} \\
      & FVT  & F         & {\tt FVTIXX}, {\tt FVTOXX} \\
      & FFT  & V         & {\tt JIOTXX} \\
      & FFV  & T         & {\tt UIOVXX} \\ \hline
 VVVT & VVVT & Amplitude & {\tt VVVTXX} \\
      & VVT  & V         & {\tt JVVTXX} \\
      & VVV  & T         & {\tt UVVVXX} \\
\hline
 VVVVT & GGGGT & Amplitude & {\tt GGGGTX} \\
       & GGGT  & G         & {\tt JGGGTX} \\
       & GGGG  & T         & {\tt UGGGGX} \\ \hline
\end{tabular}
\caption{List of the new vertex subroutines in {\tt HELAS} system.}
\label{sublist}
\end{table}

\subsection{Tensor wavefunction}\label{sec:wavefunc}

\subsubsection{\tt TXXXXX}\label{sec:txxxxx}

This subroutine computes the spin-2 {\tt T}ensor particle
wavefunction; namely, $\epsilon^{\mu\nu}(p,\lambda)$ and
$\epsilon^{\mu\nu}(p,\lambda)^*$, in terms of its four-momentum $p$
and helicity $\lambda$, and should be called as
\begin{align*}
  {\tt CALL\ TXXXXX(P,TMASS,NHEL,NST\ ,\ TC)}
\end{align*}
The input {\tt P(0:3)} is a real four-dimensional array which
contains the four-momentum $p^{\mu}$ of the tensor boson,
{\tt TMASS} is its mass, {\tt NHEL}
(${\tt =\pm 2,\pm 1,0}$) specifies its helicity $\lambda$, and {\tt NST}
specifies whether the boson is in the final state ({\tt NST = 1}) or in
the initial state ({\tt NST = -1}). The output {\tt TC(18)} is a complex
18-dimensional array, among which the first 16 components contain the
wavefunction as
\begin{align}
 {\tt TC(4\mu+\nu+1)=T(\mu+1,\nu+1)},
\label{TtoTC}
\end{align}
namely
\begin{align*}
 &{\tt TC(\ 1)=T(1,1)}, \nn \\
 &{\tt TC(\ 2)=T(1,2)}, \nn \\
 &{\tt TC(\ 3)=T(1,3)}, \nn \\
 &{\tt TC(\ 4)=T(1,4)}, \nn \\
 &{\tt ......       } \nn \\
 &{\tt TC(16)=T(4,4)},
\end{align*}
where
\begin{align}
 {\tt T(\mu+1,\nu+1)}=
 \begin{cases}
  \epsilon^{\mu\nu}(p,\lambda)^* &\text{for {\tt NST = 1}}, \\
  \epsilon^{\mu\nu}(p,\lambda)   &\text{for {\tt NST = -1}},
 \end{cases}
\label{Tmunu}
\end{align}
and the last two contain the flowing-out
four-momentum
\begin{align}
  ({\tt TC(17)},\,{\tt TC(18)})
 ={\tt NST}\,({\tt P(0)}+i{\tt P(3)},\,{\tt P(1)}+i{\tt P(2)}).
\end{align}

The helicity states of the tensor boson can be expressed as
\begin{align}
 \eps^{\mu\nu}(p,\pm 2) &= \eps^{\mu}(p,\pm)\,\eps^{\nu}(p,\pm), \nn\\
 \eps^{\mu\nu}(p,\pm 1) &= \frac{1}{\sqrt{2}}
               \big[ \eps^{\mu}(p,\pm)\,\eps^{\nu}(p,0)
                    +\eps^{\mu}(p,0)\,\eps^{\nu}(p,\pm)\big], \nn\\
 \eps^{\mu\nu}(p, 0) &= \frac{1}{\sqrt{6}}
               \big[  \eps^{\mu}(p,+)\,\eps^{\nu}(p,-)
                    + \eps^{\mu}(p,-)\,\eps^{\nu}(p,+) \nn\\
       &\qquad\quad +2\,\eps^{\mu}(p,0)\,\eps^{\nu}(p,0)\big],
\end{align}
by using the vector boson wavefunctions $\epsilon^\mu(p,\lambda)$ that
obey the relation
\begin{align}
 J_-\,\eps^{\mu}(p,\lambda)=\eps^{\mu}(p,\lambda-1),
\end{align}
where $J_-=J_x-iJ_y$ is the $J_z$ lowering operator. The spin-1 vector
wavefunction in the {\tt HELAS} convention~\cite{helas} satisfies this
relation, and hence we simply use the {\tt HELAS} code to obtain the
tensor wavefunction. These tensor wavefunctions are traceless,
transverse, orthogonal, and symmetric,
\begin{align}
  \epsilon(p,\lambda)^\mu_{\ \mu}=0,\quad
  p_\mu\epsilon^{\mu\nu}(p,\lambda)
 =p_{\nu}\epsilon^{\mu\nu}(p,\lambda)=0, \nn\\
  \epsilon^{\mu\nu}(p,\lambda)\epsilon_{\mu\nu}(p,\lambda^\prime)^*
 =\delta_{\lambda\lambda^\prime},\quad
  \epsilon^{\mu\nu}(p,\lambda)=\epsilon^{\nu\mu}(p,\lambda),
\end{align}
and the completeness relation is
\begin{align}
 \sum_{\lambda=-2}^{+2} \epsilon^{\mu\nu}(p,\lambda)
  \epsilon^{\alpha\beta}(p,\lambda)^*=B^{\mu\nu,\alpha\beta}(p)
\end{align}
with
\begin{align}
 &B^{\mu\nu,\alpha\beta}(p) \nn\\
  &\quad = \frac{1}{2}
     ( \eta^{\mu\alpha}\eta^{\nu\beta}
      +\eta^{\mu\beta}\eta^{\nu\alpha}
      -\eta^{\mu\nu}\eta^{\alpha\beta} ) \nn\\
  &\qquad-\frac{1}{2m_T^2}
     ( \eta^{\mu\alpha}{p^\nu p^\beta}+\eta^{\nu\beta}{p^\mu p^\alpha}
      +\eta^{\mu\beta}{p^\nu p^\alpha}+\eta^{\nu\alpha}{p^\mu p^\beta})
  \nn\\
  &\qquad+\frac{1}{6}
   \Big(\eta^{\mu\nu}+\frac{2}{m_T^2}p^\mu p^\nu\Big)
   \Big(\eta^{\alpha\beta}+\frac{2}{m_T^2}p^\alpha p^\beta\Big).
\end{align}
%

\subsection{SST vertex}\label{sec:vertex_i}

The {\tt SST} vertices are obtained from the interaction Lagrangian
among the tensor and two scalar bosons:
\begin{subequations}
\begin{align}
 {\cal L}_{\tt SST} = {\tt GT}\,T^{\mu\nu*}
  \Big[&-\eta_{\mu\nu}
        \big\{ (\partial^\rho S^*)^\ast (\partial_\rho S^*)
              -m^2_S S S^*
        \big\} \nn\\
       & +(\partial_\mu S^*)^\ast (\partial_\nu S^*)
        +(\partial_\nu S^*)^\ast (\partial_\mu S^*)
  \Big]
\end{align}
for the complex scalar field, or
\begin{align}
 {\cal L}_{\tt SST} = {\tt GT}\,T^{\mu\nu*}
  \Big[&-\eta_{\mu\nu}
        \Big\{ \frac{1}{2}\partial^\rho S^*\partial_\rho S^*
              -\frac{m^2_S}{2}S^*S^*
        \Big\} \nn\\
       &+\partial_\mu S^*\partial_\nu S^*
  \Big]
\end{align}
\end{subequations}
for the real scalar field. Here,
\begin{align}
 {\tt GT} = {\tt GTS} =-1/\Lambda
\label{GTS}
\end{align}
is the coupling constant.

\subsubsection{\tt SSTXXX}

This subroutine computes the amplitude of the {\tt SST} vertex from two
{\tt S}calar boson wavefunctions and a {\tt T}ensor boson wavefunction,
and should be called as
\begin{align*}
 {\tt CALL\ SSTXXX(S1,S2,TC,GT,SMASS\ ,\ VERTEX)}
\end{align*}
The inputs {\tt S1(3)} and {\tt S2(3)} are complex three-dimensional
arrays which contain the wavefunctions of the {\tt S}calar bosons,
{\tt S1(1)} and {\tt S2(1)}, and their four-momenta as
\begin{align*}
 p_1^{\mu} &= (\Re e{\tt S1(2)},\Re e{\tt S1(3)},
               \Im m{\tt S1(3)},\Im m{\tt S1(2)}), \\
 p_2^{\mu} &= (\Re e{\tt S2(2)},\Re e{\tt S2(3)},
               \Im m{\tt S2(3)},\Im m{\tt S2(2)}).
\end{align*}
The input {\tt TC(18)} is a complex 18-dimensional array which contains
the wavefunction of the {\tt T}ensor boson, and its four-momentum; see
the {\tt TXXXXX} subroutine in Sect.~\ref{sec:txxxxx}, {\tt GT} is the
coupling constant in Eq.~(\ref{GTS}) in units of GeV$^{-1}$, and
{\tt SMASS} is the scalar boson mass $m_S$ in GeV units. Note that all
the coupling constants in the latest {\tt HELAS} library are defined as
a complex number except for the 3-point and 4-point vector boson
couplings. The output {\tt VERTEX} is a complex number in units
of GeV:
\begin{align}
 {\tt VERTEX}={\tt GT}\,T^{\mu\nu}
  (m^2_S\eta_{\mu\nu}-C_{\mu\nu,\rho\sigma}p_1^\rho p_2^\sigma)\,
  {\tt S1(1)\,S2(1)},
\label{sst}
\end{align}
where we used the notation
\begin{align}
 T^{\mu\nu} &= {\tt T(\mu+1,\nu+1)} = {\tt TC(4\mu+\nu+1)}
\end{align}
in Eqs.~(\ref{TtoTC}) and (\ref{Tmunu}), and
\begin{align}
 C_{\mu\nu,\rho\sigma}=
\eta_{\mu\rho}\eta_{\nu\sigma}+\eta_{\mu\sigma}\eta_{\nu\rho}-
\eta_{\mu\nu}\eta_{\rho\sigma}.
\end{align}

\subsubsection{\tt HSTXXX}

This subroutine computes an off-shell scalar current {\tt H} made from
the interactions of a {\tt S}calar boson and a {\tt T}ensor boson by the
{\tt SST} vertex, and should be called as
\begin{align*}
 {\tt CALL\ HSTXXX(TC,SC,GT,SMASS,SWIDTH\ ,\ HST)}
\end{align*}
The inputs {\tt TC(18)} and {\tt SC(3)} are the wavefunctions and
momenta of the {\tt T}ensor and {\tt S}calar bosons, respectively, and
{\tt SWIDTH} is the scalar boson width $\Gamma_S$. The output
{\tt HST(3)} gives the off-shell scalar current multiplied by the scalar
boson propagator, which is expressed as a complex three-dimensional
array:
\begin{multline}
 {\tt HST(1)} = i{\tt GT}\,\frac{i}{q^2-m_S^2+im_S\Gamma_S}\,T^{\mu\nu} \\
   \times (m^2_S\eta_{\mu\nu}
        +C_{\mu\nu,\rho\sigma}p^\rho q^\sigma)\,{\tt SC(1)},
\label{hst}
\end{multline}
and
\begin{align}
 {\tt HST(2)} &= {\tt TC(17)}+{\tt SC(2)}, \\
 {\tt HST(3)} &= {\tt TC(18)}+{\tt SC(3)}.
\end{align}
Here the momenta $p$ and $q$ are
\begin{align*}
 p^{\mu} &= (\Re e{\tt SC(2)},\Re e{\tt SC(3)},
             \Im m{\tt SC(3)},\Im m{\tt SC(2)}), \\
 q^{\mu} &= (\Re e{\tt HST(2)},\Re e{\tt HST(3)},
             \Im m{\tt HST(3)},\Im m{\tt HST(2)}).
\end{align*}

\subsubsection{\tt USSXXX}

This subroutine computes an off-shell tensor current {\tt U} made from
two flowing-out {\tt S}calar bosons by the {\tt SST} vertex, and
should be called as
\begin{align*}
 {\tt CALL\ USSXXX(S1,S2,GT,SMASS,TMASS,TWIDTH\ ,\ USS)}
\end{align*}
The inputs {\tt TMASS} and {\tt TWIDTH} are the tensor boson mass
and width, $m_T$ and $\Gamma_T$. The output {\tt USS(18)}
gives the off-shell tensor current multiplied by the tensor boson
propagator, which is expressed as a complex 18-dimensional array:
\begin{multline}
 T^{\alpha\beta} = i{\tt GT}\,
  \frac{iB^{\mu\nu,\alpha\beta}}{q^2-m_T^2+im_T\Gamma_T} \\
  \times(m^2_S\eta_{\mu\nu}
    -C_{\mu\nu,\rho\sigma}p_1^\rho p_2^\sigma)\,{\tt S1(1)\,S2(1)},
\label{uss}
\end{multline}
where we used the notation
\begin{align}
 T^{\alpha\beta}={\tt T(\alpha+1,\beta+1)},
\end{align}
whose components are assigned into the first 16 component of {\tt USS}
as in Eq.~(\ref{TtoTC}), and
\begin{align}
 {\tt USS(17)} &= {\tt S1(2)}+{\tt S2(2)},
\label{qUSS1}\\
 {\tt USS(18)} &= {\tt S1(3)}+{\tt S2(3)}.
\label{qUSS2}
\end{align}
Here, $p_1$ and $p_2$ are the momenta of the outgoing scalars, and
$q$ is that of the off-shell tensor boson given in Eqs.~(\ref{qUSS1}) and
(\ref{qUSS2}) as
\begin{align*}
 q^{\mu} = (\Re e{\tt USS(17)},\Re e{\tt USS(18)},
            \Im m{\tt USS(18)},\Im m{\tt USS(17)}).
\end{align*}

Although the effective Lagrangian of Eq.~(\ref{IL}) does not dictates
   the off-shell behavior of the gravitons, we allow gravitons
   to propagate just once in the total amplitude where there
   are no external gravitons in the initial or final states.
   This is convenient when studying the correlated decays of
   the graviton production and its subsequent decays.

  We may also note that the order of the {\tt GT} couplings should be
  restricted to 1 when there is an external graviton,
  and 2 when a graviton is exchanged among the SM
  particles.\\

Before turning to the {\tt FFT} vertex,
it should be noticed here that the conventional factors of $i$ in the
vertices and those in the propagators are both included in the off-shell
wavefunctions, such as Eqs.~(\ref{hst}) and (\ref{uss}) above, according
to the {\tt HELAS} convention. The {\tt HELAS} amplitude, obtained by
the vertices, such as Eq.~(\ref{sst}), gives the contribution to the $T$
matrix element without the factor of $i$. See more details in the
{\tt HELAS} manual~\cite{helas}.

\subsection{FFT vertex}

The {\tt FFT} vertices are obtained from the interaction Lagrangian among
the tensor boson and two fermions:
\begin{multline}
 {\cal L}_{\tt FFT} = 4{\tt GT}\,T^{\mu\nu*}
  \Big[-\eta_{\mu\nu}
         \Big\{ \bar f(i\gamma^\rho\partial_\rho -m_F)f
               -\frac{1}{2}\partial^\rho(\bar fi\gamma_\rho f)
         \Big\} \\
  +\Big\{ \frac{1}{2}\bar fi\gamma_\mu \partial_\nu f
               -\frac{1}{4}\partial_\mu(\bar fi\gamma_\nu f)
               +(\mu\leftrightarrow\nu)
        \Big\}
  \Big],
\end{multline}
where the coupling constant is
\begin{align}
 {\tt GT} = {\tt GTF} =-1/(4\Lambda).
\label{GTF}
\end{align}

\subsubsection{\tt IOTXXX}

This subroutine computes the amplitude of the {\tt FFT} vertex from a
flowing-{\tt I}n fermion spinor, a flowing-{\tt O}ut fermion spinor and
a {\tt T}ensor boson wavefunction, and should be called as
\begin{align*}
  {\tt CALL\ IOTXXX(FI,FO,TC,GT,FMASS\ ,\ VERTEX)}
\end{align*}
The inputs {\tt FI(6)} and {\tt FO(6)} are complex six-dimensional
arrays which contain the wavefunctions of the flowing-{\tt I}n and
flowing-{\tt O}ut {\tt F}ermions, and their four-momenta as
\begin{align*}
 p_1^{\mu} &= (\Re e{\tt FI(5)},\Re e{\tt FI(6)},
               \Im m{\tt FI(6)},\Im m{\tt FI(5)}), \\
 p_2^{\mu} &= (\Re e{\tt FO(5)},\Re e{\tt FO(6)},
               \Im m{\tt FO(6)},\Im m{\tt FO(5)}).
\end{align*}
The input 
{\tt GT} is the coupling constant in Eq.~(\ref{GTF}),
and {\tt FMASS} is the fermion mass $m_F$.
What we compute here is
\begin{multline}
 {\tt VERTEX} = {\tt GT}\,{\tt (FO)}
  \big[T^{\mu\nu}\big\{-\eta_{\mu\nu}(\not{\!p}_1+\not{\!p}_2-2m_F) \\
       +\gamma_\mu(p_1+p_2)_\nu+(\mu\leftrightarrow\nu)\big\}
  \big]{\tt (FI)},
\end{multline}
where we use the notation
\begin{align}
 {\tt (FI)} &= \begin{pmatrix}\,{\tt FI(1)}\,\\\,{\tt FI(2)}\,\\
                              \,{\tt FI(3)}\,\\\,{\tt FI(4)}\,
               \end{pmatrix}, \\
 {\tt (FO)} &= ({\tt FO(1)},\,{\tt FO(2)},\,{\tt FO(3)},\,{\tt FO(4)}).
\end{align}

\subsubsection{\tt FTIXXX}

This subroutine computes an off-shell {\tt F}ermion wavefunction made
from the interactions of a {\tt T}ensor boson and a flowing-{\tt I}n
fermion by the {\tt FFT} vertex, and should be called as
\begin{align*}
 {\tt CALL\ FTIXXX(FI,TC,GT,FMASS,FWIDTH\ ,\ FTI)}
\end{align*}
The output {\tt FTI(6)} gives the off-shell fermion wavefunction
multiplied by the fermion propagator, which is expressed as a complex
six-dimensional array:
\begin{align}
 {\tt (FTI)} &= i{\tt GT}\,
   \frac{i(\not{\!q}+m_F)}{q^2-m_F^2+im_F\Gamma_F}\,T^{\mu\nu} \nn\\
 &\quad\times\big[-\eta_{\mu\nu}(\not{\!p}+\not{\!q}-2m_F) \nn\\
 &\qquad\ \ +\gamma_\mu(p+q)_\nu +(\mu\leftrightarrow\nu)
  \big]\,{\tt (FI)},
\end{align}
where we use the notation
\begin{align}
 {\tt (FTI)} = \begin{pmatrix}\,{\tt FTI(1)}\,\\\,{\tt FTI(2)}\,\\
                              \,{\tt FTI(3)}\,\\\,{\tt FTI(4)}\,
               \end{pmatrix},
\end{align}
and
\begin{align}
 {\tt FTI(5)} &= {\tt FI(5)}-{\tt TC(17)}, \\
 {\tt FTI(6)} &= {\tt FI(6)}-{\tt TC(18)}.
\end{align}
Here the momenta $p$ and $q$ are
\begin{align*}
 p^{\mu} &= (\Re e{\tt FI(5)},\Re e{\tt FI(6)},
             \Im m{\tt FI(6)},\Im m{\tt FI(5)}), \\
 q^{\mu} &= (\Re e{\tt FTI(5)},\Re e{\tt FTI(6)},
             \Im m{\tt FTI(6)},\Im m{\tt FTI(5)}).
\end{align*}

\subsubsection{\tt FTOXXX}

This subroutine computes an off-shell {\tt F}ermion wavefunction made
from the interactions of a {\tt T}ensor boson and a flowing-{\tt O}ut
fermion by the {\tt FFT} vertex, and should be called as
\begin{align*}
 {\tt CALL\ FTOXXX(FO,TC,GT,FMASS,FWIDTH\ ,\ FTO)}
\end{align*}
The output {\tt FTO(6)} is a complex six-dimensional array:
\begin{align}
 {\tt (FTO)} &= i{\tt GT}\,{\tt (FO)}\,T^{\mu\nu} \nn\\
  &\quad\times\big[-\eta_{\mu\nu}(\not{\!p}+\not{\!q}-2m_F)
                 +\gamma_\mu(p+q)_\nu +(\mu\leftrightarrow\nu)
   \big] \nn\\
  &\quad\times\frac{i(\not{\!q}+m_F)}{q^2-m_F^2+im_F\Gamma_F},
\end{align}
where we use the notation
\begin{align}
 {\tt (FTO)} = ({\tt FTO(1)},\,{\tt FTO(2)},\,{\tt FTO(3)},\,{\tt FTO(4)}),
\end{align}
and
\begin{align}
 {\tt FTO(5)} &= {\tt FO(5)}+{\tt TC(17)}, \\
 {\tt FTO(6)} &= {\tt FO(6)}+{\tt TC(18)}.
\end{align}
Here the momenta $p$ and $q$ are
\begin{align*}
 p^{\mu} &= (\Re e{\tt FO(5)},\Re e{\tt FO(6)},
             \Im m{\tt FO(6)},\Im m{\tt FO(5)}), \\
 q^{\mu} &= (\Re e{\tt FTO(5)},\Re e{\tt FTO(6)},
             \Im m{\tt FTO(6)},\Im m{\tt FTO(5)}).
\end{align*}

\subsubsection{\tt UIOXXX}

This subroutine computes the bi-spinor tensor current {\tt U} made from
flowing-{\tt I}n and flowing-{\tt O}ut fermions by the {\tt FFT} vertex,
and should be called as
\begin{align*}
 {\tt CALL\ UIOXXX(FI,FO,GT,FMASS,TMASS,TWIDTH\ ,\ UIO)}
\end{align*}
The output {\tt UIO(18)} is a complex 18-dimensional array:
\begin{align}
 T^{\alpha\beta} &= i{\tt GT}\,
  \frac{iB^{\mu\nu,\alpha\beta}}{q^2-m_T^2+im_T\Gamma_T} \nn\\
 &\quad\times({\tt FO})
    \big[-\eta_{\mu\nu}(\not{\!p}_1+\not{\!p}_2-2m_F) \nn\\
      &\quad\qquad\quad\ \, +\gamma_\mu(p_1+p_2)_\nu +(\mu\leftrightarrow\nu)
    \big]({\tt FI})
\end{align}
for the first 16 components of {\tt UIO}, and
\begin{align}
 {\tt UIO(17)} &= -{\tt FI(5)}+{\tt FO(5)}, \\
 {\tt UIO(18)} &= -{\tt FI(6)}+{\tt FO(6)}.
\end{align}
Here, $p_1$ and $p_2$ are the momenta of the flowing-in and flowing-out
fermions, respectively, and $q$ is that of the tensor particle.

\subsection{VVT vertex}

The {\tt VVT} vertices are obtained from the interaction Lagrangian among
the tensor and two vector bosons:
\begin{subequations}
\begin{align}
 {\cal L}_{\tt VVT} = {\tt GT}\, T^{\mu\nu*}
    \Big[& \eta_{\mu\nu}
            \Big\{ \frac{1}{2}( \partial^\rho V^{\sigma*}
                               -\partial^\sigma V^{\rho*})^*
                              ( \partial_\rho V^{\ast}_{\sigma}
                               -\partial_\sigma V^{\ast}_{\rho}) \nn\\
  &\qquad -m_V^2V^{\rho} V^\ast_{\rho}
            \Big\} \nn\\
  &-\big\{(\partial_\mu V^{\rho*}-\partial^\rho V^{\ast}_{\mu})^*
          (\partial_\nu V^{\ast}_{\rho}-\partial_\rho V^{\ast}_{\nu}) \nn\\
  &\quad\ +m^2_VV_{\mu} V^\ast_{\nu}+(\mu\leftrightarrow\nu)\big\}
    \Big]
\end{align}
for the complex vector bosons, or $W$ bosons, and
\begin{align}
 {\cal L}_{\tt VVT} &= {\tt GT}\, T^{\mu\nu*}
    \Big[ \eta_{\mu\nu}
            \Big\{ \frac{1}{4}( \partial^\rho V^{\sigma*}
                               -\partial^\sigma V^{\rho*})
                              ( \partial_\rho V^{\ast}_{\sigma}
                               -\partial_\sigma V^{\ast}_{\rho}) \nn\\
  &\qquad\qquad\qquad\ -\frac{m_V^2}{2}V^{\rho*} V^\ast_{\rho}
            \Big\} \nn\\
  &\quad -(\partial_\mu V^{\rho*}-\partial^\rho V^{\ast}_{\mu})
          (\partial_\nu V^{\ast}_{\rho}-\partial_\rho V^{\ast}_{\nu})
         +m^2_VV^\ast_{\mu} V^\ast_{\nu} \nn\\
  &\quad -\xi^{-1}\eta_{\mu\nu}
      \Big( \partial^\rho\partial^\sigma V^\ast_{\sigma}V^\ast_{\rho}
 +\frac{1}{2}\partial^\rho V^\ast_{\rho}\partial^\sigma V^\ast_{\sigma}
      \Big) \nn \\
  &\quad +\xi^{-1}
      \Big( \partial_\mu\partial^\rho V^\ast_{\rho}V^\ast_{\nu}
           +\partial_\nu \partial^\rho V^\ast_{\rho}V^\ast_{\mu}
      \Big)
    \Big],
\end{align}
\end{subequations}
for the real ones, or gluons, photons, and $Z$ bosons.
Here the coupling constant is
\begin{align}
 {\tt GT} = {\tt GTV} =-1/\Lambda.
\label{GTV}
\end{align}
The $\xi$ terms are the gauge-fixing terms, which vanish for massive
vector bosons in the unitary gauge. For massless vector bosons we take
$\xi=1$ in the Feynman gauge.

\subsubsection{\tt VVTXXX}

This subroutine computes the amplitude of the {\tt VVT} vertex from two
{\tt V}ector boson polarization vectors and a {\tt T}ensor wavefunction,
and should be called as
\begin{align*}
 {\tt CALL\ VVTXXX(V1,V2,TC,GT,VMASS\ ,\ VERTEX)}
\end{align*}
The inputs {\tt V1(6)} and {\tt V2(6)} are complex six-dimensional
arrays which contain the {\tt V}ector boson wavefunctions, and
their momenta as
\begin{align*}
 p_1^{\mu} &= (\Re e{\tt V1(5)},\Re e{\tt V1(6)},
               \Im m{\tt V1(6)},\Im m{\tt V1(5)}), \\
 p_2^{\mu} &= (\Re e{\tt V2(5)},\Re e{\tt V2(6)},
               \Im m{\tt V2(6)},\Im m{\tt V2(5)}).
\end{align*}
The input {\tt GT} is the coupling constant in Eq.~(\ref{GTV}), and
{\tt VMASS} is the vector boson mass $m_V$.
What we compute here is
\begin{align}
 {\tt VERTEX} = {\tt GT}\,T^{\mu\nu}
   \big[&(m^2_V+p_1\cdot p_2)\,C_{\mu\nu,\rho\sigma}
  +D_{\mu\nu,\rho\sigma}(p_1,p_2) \nn\\
  &+\xi^{-1}E_{\mu\nu,\rho\sigma}(p_1,p_2)\big]V_1^{\rho}V_2^{\sigma},
\end{align}
where we use the notation
\begin{align}
 V_1^{\mu} &= {\tt V1(\mu+1)}, \\
 V_2^{\mu} &= {\tt V2(\mu+1)},
\end{align}
and
\begin{align}
 D_{\mu\nu,\rho\sigma}(p_1,p_2) &=\eta_{\mu\nu}p_{1\sigma}p_{2\rho}
  -\big[ \eta_{\mu\sigma}p_{1\nu}p_{2\rho}
        +\eta_{\mu\rho}p_{1\sigma}p_{2\nu} \nn\\
       &\quad -\eta_{\rho\sigma}p_{1\mu}p_{2\nu}
              +(\mu\leftrightarrow\nu)\big], \\
 E_{\mu\nu,\rho\sigma}(p_1, p_2)&=\eta_{\mu\nu}(p_{1\rho}p_{1\sigma}
   +p_{2\rho}p_{2\sigma}+p_{1\rho}p_{2\sigma}) \nn \\
 &\quad -\big[\eta_{\nu\sigma}p_{1\mu}p_{1\rho}
   +\eta_{\nu\rho}p_{2\mu}p_{2\sigma}+(\mu\leftrightarrow\nu)\big].
\end{align}

\subsubsection{\tt JVTXXX}

This subroutine computes an off-shell vector current {\tt J} made from
the interactions of a {\tt V}ector boson and a {\tt T}ensor boson by the
{\tt VVT} vertex, and should be called as
\begin{align*}
 {\tt CALL\ JVTXXX(VC,TC,GT,VMASS,VWIDTH\ ,\ JVT)}
\end{align*}
The input {\tt VC(6)} is the wavefunction and momentum of the
{\tt V}ector boson.
The output
{\tt JVT(6)} gives the off-shell vector current multiplied by the vector
boson propagator, which is expressed as a complex six-dimensional array:
\begin{align}
 &{\tt JVT(\alpha+1)} = i{\tt GT}\,
  \frac{i}{q^2-m^2_V+im_V\Gamma_V}
  \Big(-\eta^{\sigma\alpha}+ \frac{q^\sigma q^\alpha}{m^2_V}\Big) \nn\\
    &\qquad\qquad\ \times T^{\mu\nu}
  \big[ (m^2_V-p\cdot q)\,C_{\mu\nu,\rho\sigma}
             +D_{\mu\nu,\rho\sigma}(p,-q)\big]V^{\rho}
\end{align}
for the massive vector boson, or
\begin{align}
 {\tt JVT(\alpha+1)} &= i{\tt GT}\,\frac{-i}{q^2}\,\eta^{\sigma\alpha}
  T^{\mu\nu}
  \big[-(p\cdot q)\,C_{\mu\nu,\rho\sigma} \nn\\
  &\quad+D_{\mu\nu,\rho\sigma}(p,-q)
              +E_{\mu\nu,\rho\sigma}(p,-q)\big]V^{\rho}
\end{align}
for the massless vector boson, and
\begin{align}
 {\tt JVT(5)} &= {\tt VC(5)}+{\tt TC(17)}, \\
 {\tt JVT(6)} &= {\tt VC(6)}+{\tt TC(18)}.
\end{align}
Here the momenta $p$ and $q$ are
\begin{align*}
 p^{\mu} &= (\Re e{\tt VC(5)},\Re e{\tt VC(6)},
             \Im m{\tt VC(6)},\Im m{\tt VC(5)}), \\
 q^{\mu} &= (\Re e{\tt JVT(5)},\Re e{\tt JVT(6)},
             \Im m{\tt JVT(6)},\Im m{\tt JVT(5)}).
\end{align*}

\subsubsection{\tt UVVXXX}

This subroutine computes an off-shell tensor current {\tt U}
made from two flowing-out {\tt V}ector bosons by the {\tt VVT}
vertex, and should be called as
\begin{align*}
 {\tt CALL\ UVVXXX(V1,V2,GT,VMASS,TMASS,TWIDTH\ ,\ UVV)}
\end{align*}
The output {\tt UVV(18)} is a complex 18-dimensional array:
\begin{align}
 T^{\alpha\beta} &= i{\tt GT}\,
   \frac{iB^{\mu\nu,\alpha\beta}}{q^2-m^2_T+im_T\Gamma_T} \nn\\
  &\quad\times\big[(m^2_V+p_1\cdot p_2)\,C_{\mu\nu,\rho\sigma}
  +D_{\mu\nu,\rho\sigma}(p_1,p_2) \nn\\
  &\quad\quad\ +\xi^{-1}E_{\mu\nu,\rho\sigma}(p_1,p_2)\big]
             V_1^{\rho}V_2^{\sigma}
\end{align}
for the first 16 components of {\tt UVV}, and
\begin{align}
 {\tt UVV(17)} &= {\tt V1(5)}+{\tt V2(5)}, \\
 {\tt UVV(18)} &= {\tt V1(6)}+{\tt V2(6)}.
\end{align}
Here, $p_1$ and $p_2$ are the momenta of the
outgoing vector bosons, and $q$ is that of the tensor boson.

\subsection{FFVT vertex}

The {\tt FFVT} vertices are obtained from the interaction Lagrangian among
the tensor boson, vector boson and two fermions:
\begin{multline}
 {\cal L}_{\tt FFVT} = -2{\tt GT}\, T^{\mu\nu*}
  (\eta_{\mu\nu}\eta_{\rho\sigma}-\eta_{\mu\sigma}\eta_{\nu\rho}) \\
 \times \bar f\gamma^\sigma
  \big[{\tt GC(1)}P_L+{\tt GC(2)}P_R\big]f\,V^{\rho*}
\end{multline}
with the chiral-projection operator $P_{R,L}=\frac{1}{2}(1\pm\gamma_5)$.
The coupling constant {\tt GT} is
\begin{align}
 {\tt GT} = {\tt GTFV} =-1/(2\Lambda),
\label{GTFV}
\end{align}
and {\tt GC(1)} and {\tt GC(2)} are the relevant {\tt FFV} left and
right coupling constants. The list of the coupling constants is shown in
Table~\ref{couplist}.
For instance, in the case of the interaction with a gluon, the {\tt FFV}
couplings are
\begin{align}
 {\tt GC(1)} = {\tt GC(2)} = -g_s,
\end{align}
where the sign of the coupling is fixed by the {\tt HELAS}
convention~\cite{helas}.

\subsubsection{\tt IOVTXX}

This subroutine computes the amplitude of the {\tt FFVT} vertex from a
flowing-{\tt I}n fermion spinor, a flowing-{\tt O}ut fermion spinor, a
{\tt V}ector boson polarization vector and a {\tt T}ensor boson
wavefunction, and should be called as
\begin{align*}
  {\tt CALL\ IOVTXX(FI,FO,VC,TC,GC,GT\ ,\ VERTEX)}
\end{align*}
What we compute here is
\begin{multline}
 {\tt VERTEX} =-\,{\tt GT}\,{\tt (FO)}
  \big[ T^{\mu\nu}
  (\eta_{\mu\nu}\eta_{\rho\sigma}-C_{\mu\nu,\rho\sigma}) \\
 \times V^{\rho}\gamma^{\sigma}({\tt GC(1)}P_L+{\tt GC(2)}P_R)\big]{\tt (FI)}.
\end{multline}

\subsubsection{\tt FVTIXX}

This subroutine computes an off-shell fermion wavefunction by the
{\tt FFVT} vertex, and should be called as
\begin{align*}
 {\tt CALL\ FVTIXX(FI,VC,TC,GC,GT,FMASS,FWIDTH\ ,\ FVTI)}
\end{align*}
What we compute here is
\begin{multline}
 {\tt (FVTI)} = -i{\tt GT}\,
  \frac{i(\not{\!q}+m_F)}{q^2-m^2_F+im_F\Gamma_F}\,T^{\mu\nu}
   (\eta_{\mu\nu}\eta_{\rho\sigma}-C_{\mu\nu,\rho\sigma}) \\
 \times V^{\rho}\gamma^{\sigma}
 \big[{\tt GC(1)}P_L+{\tt GC(2)}P_R\big]{\tt (FI)}
\end{multline}
for the first 4 components of {\tt FVTI(6)}, and
\begin{align}
 {\tt FVTI(5)} &= {\tt FI(5)}-{\tt VC(5)}-{\tt TC(17)}, \\
 {\tt FVTI(6)} &= {\tt FI(6)}-{\tt VC(6)}-{\tt TC(18)}.
\end{align}
Here the momentum $q$ is
\begin{align*}
 q^{\mu} = (\Re e{\tt FVTI(5)},\Re e{\tt FVTI(6)},
            \Im m{\tt FVTI(6)},\Im m{\tt FVTI(5)}).
\end{align*}

\subsubsection{\tt FVTOXX}

This subroutine computes an off-shell fermion wavefunction by the
{\tt FFVT} vertex, and should be called as
\begin{align*}
 {\tt CALL\ FVTOXX(FO,VC,TC,GC,GT,FMASS,FWIDTH\ ,\ FVTO)}
\end{align*}
What we compute here is
\begin{align}
 {\tt (FVTO)} &= -i{\tt GT}\,{\tt (FO)}\,T^{\mu\nu}
 (\eta_{\mu\nu}\eta_{\rho\sigma}-C_{\mu\nu,\rho\sigma}) \nn\\
  &\times V^{\rho}\gamma^{\sigma}\big[{\tt GC(1)}P_L+{\tt GC(2)}P_R\big]
  \frac{i(\not{\!q}+m_F)}{q^2-m^2_F+im_F\Gamma_F}
\end{align}
for the first 4 components of {\tt FVTO(6)}, and
\begin{align}
 {\tt FVTO(5)} &= {\tt FO(5)}+{\tt VC(5)}+{\tt TC(17)}, \\
 {\tt FVTO(6)} &= {\tt FO(6)}+{\tt VC(6)}+{\tt TC(18)}.
\end{align}
Here the momentum $q$ is
\begin{align*}
 q^{\mu} = (\Re e{\tt FVTO(5)},\Re e{\tt FVTO(6)},
            \Im m{\tt FVTO(6)},\Im m{\tt FVTO(5)}).
\end{align*}

\subsubsection{\tt JIOTXX}

This subroutine computes an off-shell vector current by the {\tt FFVT}
vertex, and should be called as
\begin{align*}
 {\tt CALL\ JIOTXX(FI,FO,TC,GC,GT,VMASS,VWIDTH\ ,\ JIOT)}
\end{align*}
What we compute here is
\begin{align}
 {\tt JIOT(\alpha+1)} &= -i{\tt GT}\,\frac{i}{q^2-m^2_V+im_V\Gamma_V}
  \Big(-\eta^{\rho\alpha}+\frac{q^\rho q^\alpha}{m^2_V}\Big) \nn\\
 &\times T^{\mu\nu}
        (\eta_{\mu\nu}\eta_{\rho\sigma}-C_{\mu\nu,\rho\sigma}) \nn\\
 &\times {\tt (FO)} \big[\gamma^{\sigma}
 \{{\tt GC(1)}P_L+{\tt GC(2)}P_R\}\big]{\tt (FI)}
\end{align}
for the massive vector boson, or
\begin{multline}
 {\tt JIOT(\alpha+1)} = -i{\tt GT}\,\frac{-i}{q^2}\,\eta^{\rho\alpha}\,
  T^{\mu\nu}
        (\eta_{\mu\nu}\eta_{\rho\sigma}-C_{\mu\nu,\rho\sigma}) \\
  \times {\tt (FO)} \big[\gamma^{\sigma}
  \{{\tt GC(1)}P_L+{\tt GC(2)}P_R\}\big]{\tt (FI)}
\end{multline}
for the massless vector boson, and
\begin{align}
 {\tt JIOT(5)} &= -{\tt FI(5)}+{\tt FO(5)}+{\tt TC(17)}, \\
 {\tt JIOT(6)} &= -{\tt FI(6)}+{\tt FO(6)}+{\tt TC(18)}.
\end{align}
Here $q$ is the momentum of the vector boson.

\subsubsection{\tt UIOVXX}

This subroutine computes an off-shell tensor current by the {\tt FFVT}
vertex, and should be called as
\begin{align*}
 {\tt CALL\ UIOVXX(FI,FO,VC,GC,GT,TMASS,TWIDTH\ ,\ UIOV)}
\end{align*}
What we compute here is
\begin{multline}
 T^{\alpha\beta} = -i{\tt GT}\,
    \frac{iB^{\mu\nu,\alpha\beta}}{q^2-m^2_T+im_T\Gamma_T}\,
 (\eta_{\mu\nu}\eta_{\rho\sigma}-C_{\mu\nu,\rho\sigma}) \\
  \times {\tt (FO)}\big[V^{\rho}\gamma^{\sigma}
   \{{\tt GC(1)}P_L+{\tt GC(2)}P_R\}\big]({\tt FI})
\end{multline}
for the first 16 components of {\tt UIOV(18)}, and
\begin{align}
 {\tt UIOV(17)} &= -{\tt FI(5)}+{\tt FO(5)}+{\tt VC(5)}, \\
 {\tt UIOV(18)} &= -{\tt FI(6)}+{\tt FO(6)}+{\tt VC(6)}.
\end{align}
Here $q$ is the momentum of the tensor boson.

\subsection{VVVT vertex}

The {\tt VVVT} vertices are obtained from the interaction Lagrangian
among the tensor and three vector bosons:
\begin{align}
 {\cal L}_{\tt VVVT} &= {\tt GT\,GC}\,f^{abc}\, T^{\mu\nu*}
  \Big[\frac{1}{2}\eta_{\mu\nu}
         (\partial_\rho V^a_\sigma -\partial_\sigma V^a_\rho )
         V^{b,\rho}V^{c,\sigma} \nn\\
       &\ -(\partial_{\mu} V^{a,\rho}-\partial^{\rho}V^{a}_\mu)
         V^{b}_\nu V^{c}_\rho
       -(\partial_{\nu} V^{a}_\rho-\partial_{\rho}V^{a}_\nu)
         V^{b,\rho} V^{c}_\mu
  \Big]
\label{L_VVVT}
\end{align}
with the structure constant $f^{abc}$ and
the coupling constant, as in Eq.~(\ref{GTV}),
\begin{align}
 {\tt GT} = {\tt GTV} = -1/\Lambda.
\end{align}

In this paper we concentrate on the interactions with gluons for the
{\tt VVVT} vertex, so in this case {\tt GC} is the strong coupling
constant, 
\begin{align}
 {\tt GC} =g_s,
\end{align}
and $f^{abc}$ is the structure constants of the group $SU(3)$, which can 
be handled by the {\tt MG} automatically. 
As in the original subroutines for the {\tt VVV} vertex~\cite{helas},
the following subroutines, {\tt VVVTXX}, {\tt JVVTXX}, and {\tt UVVVXX},
can be used for the electroweak gauge bosons without any modifications.

\subsubsection{\tt VVVTXX}

This subroutine computes the amplitude of the {\tt VVVT} vertex from
three {\tt V}ector boson polarization vectors and a {\tt T}ensor boson
wavefunction, and should be called as
\begin{align*}
 {\tt CALL\ VVVTXXX(VA,VB,VC,TC,GC,GT\ ,\ VERTEX)}
\end{align*}
What we compute here is
\begin{align}
 {\tt VERTEX} = &-{\tt GT\,GC}\,T^{\mu\nu}
   \big[ C_{\mu\nu,\rho\sigma}(p_{a\lambda}-p_{b\lambda}) \nn\\
 &\qquad\quad +C_{\mu\nu,\sigma\lambda}(p_{b\rho}-p_{c\rho})
  +C_{\mu\nu,\lambda\rho}(p_{c\sigma}-p_{a\sigma}) \nn\\
 &\qquad\quad +F_{\mu\nu,\rho\sigma\lambda}(p_a,p_b,p_c)\big]
  V_a^{\rho}V_b^{\sigma}V_c^{\lambda}
\end{align}
with
\begin{multline}
 F_{\mu\nu,\rho\sigma\lambda} (p_a,p_b,p_c)
  =\eta_{\mu\rho}\eta_{\sigma\lambda}(p_b-p_c)_\nu
+\eta_{\mu\sigma}\eta_{\rho\lambda}(p_c-p_a)_\nu \\
  +\eta_{\mu\lambda}\eta_{\rho\sigma}(p_a-p_b)_\nu
+ (\mu\leftrightarrow\nu).
\end{multline}
Here, the vector bosons (gluons in this paper) {\tt VA}, {\tt VB},
and {\tt VC} have the momentum $p_a$, $p_b$, and $p_c$, and the color
$a$, $b$, and $c$, respectively.

\subsubsection{\tt JVVTXX}

This subroutine computes an off-shell vector current by the {\tt VVVT}
vertex, and should be called as
\begin{align*}
 {\tt CALL\ JVVTXX(VA,VB,TC,GC,GT,VMASS,VWIDTH\ ,\ JVVT)}
\end{align*}
What we compute here is
\begin{align}
  &{\tt JVVT(\alpha+1)} \nn\\ &\quad = -i{\tt GT\,GC}\,
\frac{i}{q^2-m^2_V+im_V\Gamma_V}
  \Big(-\eta^{\lambda\alpha}+\frac{q^\lambda q^\alpha}{m^2_V}\Big) \nn\\ 
   & \qquad \times T^{\mu\nu}
   \big[ C_{\mu\nu,\rho\sigma}(p_{a\lambda}-p_{b\lambda})
        +C_{\mu\nu,\sigma\lambda}(p_{b\rho}+q_{\rho}) \nn\\
        &\qquad\quad +C_{\mu\nu,\lambda\rho}(-q_{\sigma}-p_{a\sigma})
  +F_{\mu\nu,\rho\sigma\lambda}(p_a,p_b,-q)\big]
  V_a^{\rho}V_b^{\sigma}
\end{align}
for the massive vector boson, or
\begin{align}
  {\tt JVVT(\alpha+1)} = -&i{\tt GT\,GC}\,
   \frac{-i}{q^2}\,\eta^{\lambda\alpha}T^{\mu\nu}
   \big[ C_{\mu\nu,\rho\sigma}(p_{a\lambda}-p_{b\lambda}) \nn\\
  &+C_{\mu\nu,\sigma\lambda}(p_{b\rho}+q_{\rho})
        +C_{\mu\nu,\lambda\rho}(-q_{\sigma}-p_{a\sigma}) \nn\\
  &+F_{\mu\nu,\rho\sigma\lambda}(p_a,p_b,-q)\big]
  V_a^{\rho}V_b^{\sigma},
\end{align}
for the massless vector boson, and
\begin{align}
 {\tt JVVT(5)} &= {\tt VA(5)}+{\tt VB(5)}+{\tt TC(17)}, \\
 {\tt JVVT(6)} &= {\tt VA(6)}+{\tt VB(6)}+{\tt TC(18)}.
\end{align}
Here the momenta $p_a$, $p_b$ and $q$ are
\begin{align*}
 p_a^{\mu} &= (\Re e{\tt VA(5)},\Re e{\tt VA(6)},
             \Im m{\tt VA(6)},\Im m{\tt VA(5)}), \\
 p_b^{\mu} &= (\Re e{\tt VB(5)},\Re e{\tt VB(6)},
             \Im m{\tt VB(6)},\Im m{\tt VB(5)}), \\
 q^{\mu} &= (\Re e{\tt JVVT(5)},\Re e{\tt JVVT(6)},
             \Im m{\tt JVVT(6)},\Im m{\tt JVVT(5)}).
\end{align*}
Note that the off-shell gluon {\tt JVVT} has the color $c$.

\subsubsection{\tt UVVVXX}

This subroutine computes an off-shell tensor current by the {\tt VVVT}
vertex, and should be called as
\begin{align*}
 {\tt CALL\ UVVVXX(VA,VB,VC,GC,GT,TMASS,TWIDTH\ ,\ UVVV)}
\end{align*}
What we compute here is
\begin{align}
  T^{\alpha\beta} = -i{\tt GT\,GC}\,
   &\frac{iB^{\mu\nu,\alpha\beta}}{q^2-m^2_T+im_T\Gamma_T}
   \big[ C_{\mu\nu,\rho\sigma}(p_{a\lambda}-p_{b\lambda}) \nn\\
  &\ \ +C_{\mu\nu,\sigma\lambda}(p_{b\rho}-p_{c\rho})
         +C_{\mu\nu,\lambda\rho}(p_{c\sigma}-p_{a\sigma}) \nn\\
  &\ \ +F_{\mu\nu,\rho\sigma\lambda}(p_a,p_b,p_c)\big]
  V_a^{\rho}V_b^{\sigma}V_c^{\lambda}
\end{align}
for the first 16 components of {\tt UVVV(18)}, and
\begin{align}
 {\tt UVVV(17)} &= {\tt VA(5)}+{\tt VB(5)}+{\tt VC(5)}, \\
 {\tt UVVV(18)} &= {\tt VA(6)}+{\tt VB(6)}+{\tt VC(6)}.
\end{align}
Here $p_a$, $p_b$ and $p_c$ are the momenta of the outgoing
vector bosons, and $q$ is that of the tensor boson.

\subsection{VVVVT vertex}\label{sec:vertex_f}

The {\tt VVVVT} vertices are obtained from the interaction
Lagrangian among the tensor and four vector bosons:
\begin{align}
 {\cal L}_{\tt VVVVT} 
  = &-{\tt GT\,GC}^2\, f^{abe}f^{cde}\, T^{\mu\nu*} \nn\\
  &\ \,\times\Big[\frac{1}{4}\eta_{\mu\nu}
   V^{a,\rho*}V^{b,\sigma*}V^{c\ast}_\rho V^{d\ast}_\sigma
  -V^{b,\rho*}V^{a*}_\mu V^{c*}_{\nu}V^{d*}_\rho\Big]
\end{align}
with the coupling constants {\tt GT} = {\tt GTV} in Eq.~(\ref{GTV}) and
{\tt GC} = $g_s$ for the interactions with gluons.

We should note that the 5-point vertex cannot be generated by {\tt MG},
and thus we must add the following subroutines, {\tt GGGGTX},
{\tt JGGGTX}, or {\tt UGGGGX}, to the amplitudes which have the
corresponding color structures by hand.

\subsubsection{\tt GGGGTX}

This subroutine computes the portion of the amplitude of the 
{\tt VVVVT} vertex from four {\tt G}luon polarization vectors and a 
{\tt T}ensor boson wavefunction corresponding to the color structure
$f^{abe}f^{cde}$, and should be called as
\begin{align*}
 {\tt CALL\ GGGGTX(VA,VB,VC,VD,TC,GC,GT\ ,\ VERTEX)}
\end{align*}
The output is
\begin{align}
 {\tt VERTEX}=-{\tt GT\,GC}^2\,T^{\mu\nu}\,
  G_{\mu\nu,\rho\lambda\sigma\delta}\,
 V_a^{\rho}V_b^{\sigma}V_c^{\lambda}V_d^{\delta}
\end{align}
with
\begin{align}
 G_{\mu\nu,\rho\sigma\lambda\delta}
  &=\eta_{\mu\nu}( \eta_{\rho\sigma}\eta_{\lambda\delta}
                  -\eta_{\rho\delta}\eta_{\sigma\lambda}) \nn\\
  &\quad+\big[ \eta_{\mu\rho}\eta_{\nu\delta}\eta_{\lambda\sigma}
            +\eta_{\mu\lambda}\eta_{\nu\sigma}\eta_{\rho\delta}
            -\eta_{\mu\rho}\eta_{\nu\sigma}\eta_{\lambda\delta} \nn\\
  &\qquad\ -\eta_{\mu\lambda}\eta_{\nu\delta}\eta_{\rho\sigma}
  +(\mu\leftrightarrow\nu)\big].
\end{align}

To obtain the complete amplitude, this subroutine must be called
three times (once for each color structure) with the following
permutations:
\begin{align*}
  {\tt CALL\ GGGGTX(VA,VB,VC,VD,TC,GC,GT\ ,\ VERTEX1)} \\[-1mm]
  {\tt CALL\ GGGGTX(VA,VC,VD,VB,TC,GC,GT\ ,\ VERTEX2)} \\[-1mm]
  {\tt CALL\ GGGGTX(VA,VD,VB,VC,TC,GC,GT\ ,\ VERTEX3)}
\end{align*}
corresponding to the color structure $f^{abe}f^{cde}$,
$f^{ace}f^{dbe}$, and $f^{ade}f^{bce}$, respectively.

\subsubsection{\tt JGGGTX}

This subroutine computes the portion of the off-shell gluon current
by the {\tt VVVVT} vertex, corresponding to the color structure
$f^{abe}f^{cde}$, and should be called as
\begin{align*}
 {\tt CALL\ JGGGTX(VA,VB,VC,TC,GC,GT\ ,\ JGGGT)}
\end{align*}
What we compute here is
\begin{align}
 {\tt JGGGT(\alpha+1)} =-i{\tt GT\,GC}^2\,\frac{-i}{q^2}\,
   \eta^{\delta\alpha}\,T^{\mu\nu}\,
   G_{\mu\nu,\rho\lambda\sigma\delta}\,
   V_a^{\rho}V_b^{\sigma}V_c^{\lambda},
\end{align}
and
\begin{align}
 {\tt JGGGT(5)} &= {\tt VA(5)}+{\tt VB(5)}+{\tt VC(5)}+{\tt TC(17)}, \\
 {\tt JGGGT(6)} &= {\tt VA(6)}+{\tt VB(6)}+{\tt VC(6)}+{\tt TC(18)}.
\end{align}
Note that the off-shell gluon {\tt JGGGT} has the color $d$ and the
momentum $q$.

\subsubsection{\tt UGGGGX}

This subroutine computes the portion of the off-shell tensor
current by the {\tt VVVVT} vertex, corresponding to the color structure
$f^{abe}f^{cde}$, and should be called as
\begin{align*}
 {\tt CALL\ UGGGGX(VA,VB,VC,VD,GC,GT,TMASS,TWIDTH\, ,\, UGGGG)}
\end{align*}
What we compute here is
\begin{align}
 T^{\alpha\beta} &= -i{\tt GT\,GC}^2
  \frac{iB^{\mu\nu,\alpha\beta}}{q^2-m^2_T+im_T\Gamma_T}\,
  G_{\mu\nu,\rho\lambda\sigma\delta}\,
  V_a^{\rho}V_b^{\sigma}V_c^{\lambda}V_d^{\delta}
\end{align}
for the first 16 components of {\tt UGGGG(18)}, and
\begin{align}
 {\tt UGGGG(17)} &= {\tt VA(5)}+{\tt VB(5)}+{\tt VC(5)}+{\tt VD(5)}, \\
 {\tt UGGGG(18)} &= {\tt VA(6)}+{\tt VB(6)}+{\tt VC(6)}+{\tt VD(6)}.
\end{align}
Here $q$ is the momentum of the tensor boson.

\subsection{Checking for the new HELAS subroutines}\label{sec:check}

The new {\tt HELAS} subroutines are tested by using the QCD gauge
invariance and the general coordinate transformation invariance of the
helicity amplitudes. In particular, we use the following processes; 
\begin{align}
 & q\bar{q}\rightarrow gT
  &&{\rm for}\ {\tt IOTXXX}, {\tt FTIXXX}, {\tt FTOXXX}, {\tt IOVTXX}, \nn\\
 & gg\rightarrow gT
  &&{\rm for}\ {\tt VVTXXX}, {\tt JVTXXX}, {\tt VVVTXX}, \nn\\
 & q\bar{q}\rightarrow q\bar{q}(gg)T
  &&{\rm for}\ {\tt FVTIXX}, {\tt FVTOXX}, {\tt JIOTXX}, \nn\\
 & gg\rightarrow ggT
  &&{\rm for}\ {\tt JVVTXX}, {\tt GGGGTX}, \nn\\
 & q\bar{q}(gg)\rightarrow T\rightarrow gg
  &&{\rm for}\ {\tt UIOXXX}, {\tt UVVXXX}. \nn
\end{align}
More explicitly, we express the helicity amplitudes of the above
processes as 
\begin{align}
 {\cal M}_{\lambda_T\lambda_g} = T_{\mu\nu\rho}\,
      \epsilon^{\mu\nu}(p_T,\lambda_T)^\ast\,
      \epsilon^\rho(p_g,\lambda_g)^*
\end{align}
with an external tensor and a gluon wavefunction.
The identities,
\begin{align}
   p_T^\mu/p_T^0\, T_{\mu\nu\rho}\, \epsilon^\rho(p_g,\lambda_g)^*
 = p_T^\nu/p_T^0\, T_{\mu\nu\rho}\, \epsilon^\rho(p_g,\lambda_g)^*
 = 0
\end{align}
for the general coordinate transformation symmetry and
\begin{align}
    p_g^\rho/p_g^0\, T_{\mu\nu\rho} = 0
\end{align}
for the SU(3) gauge invariance, test all the above subroutines thoroughly.

The subroutines {\tt UIOVXX}, {\tt UVVVXX}, {\tt JGGGTX} and {\tt UGGGGX}
have been checked in such a way that the above subroutines are used
rewriting the helicity amplitudes of the processes 
\begin{align}
 & q\bar{q}\rightarrow T \rightarrow q\bar{q}g(ggg), \nn\\
 & gg\rightarrow ggT \rightarrow ggZZ, \nn\\
 & gg\rightarrow ggggT. \nn
\end{align}
We test the agreement of the amplitudes between the expressions that
uses the above subroutines and those without them for all helicity
combinations and at arbitrary Lorentz frame. 

\section{MadGraph/MadEvent implementing for spin-2 gravitons}
\label{sec:mgme}

In this section, we would like to describe how we implement spin-2
gravitons into {\tt MG/ME}.

First, using the User Model framework in {\tt MG}~\cite{newmad}, we make
our new model directories for both the ADD and RS models, including the
massive gravitons ({\tt particles.dat}) and their interactions with the
SM particles ({\tt couplings.f} and {\tt interactions.dat}); see also
Table~\ref{couplist}. Then we insert all the new {\tt HELAS} subroutines
for spin-2 tensor bosons into the {\tt HELAS} library in {\tt MG}. Since
the present {\tt MG} does not handle external spin-2 particles, we
further modify the codes in {\tt MG} to tell it how to generate the 
{\tt SST}, {\tt FFT} and {\tt FFVT} type of vertices and helicity
amplitudes (for {\tt VVT} and {\tt VVVT} type, it has already been done
for the Higgs effective field theory ({\tt HEFT}) model), and how to
deal with the helicity of the spin-2 tensor bosons when they are
external. Moreover, since {\tt MG} can only generate Feynman diagrams
with up to 4-point vertices, the amplitudes and their {\tt HELAS} codes
with the 5-point vertex, {\tt GGGGTX}, {\tt JGGGTX},
or {\tt UGGGGX}, have been added by hand; 
see more details in Sect.~\ref{sec:vertex_f}.

\begin{table}
\centering
\begin{tabular}{cccccccccc}
\hline\hline
\multicolumn{8}{l}{3-point couplings}  & {\tt GT}  & \\
\hline
 SST &&&& {\tt h} & {\tt h} & {\tt y} && {\tt GTS} & \\
 FFT &&&& {\tt f} & {\tt f} & {\tt y} && {\tt GTF} & \\
 VVT &&&& {\tt v} & {\tt v} & {\tt y} && {\tt GTV} & \\
\hline\hline
\multicolumn{8}{l}{4-point couplings}        &{\tt GT}   &{\tt GC}  \\
\hline
 FFVT &&&{\tt d} &{\tt d} &{\tt a} &{\tt y} &&{\tt GTFV} &{\tt GAD} \\
      &&&{\tt u} &{\tt u} &{\tt a} &{\tt y} &&{\tt GTFV} &{\tt GAU} \\
      &&&{\tt l} &{\tt l} &{\tt a} &{\tt y} &&{\tt GTFV} &{\tt GAL} \\
      &&&{\tt d} &{\tt d} &{\tt z} &{\tt y} &&{\tt GTFV} &{\tt GZD} \\
      &&&{\tt u} &{\tt u} &{\tt z} &{\tt y} &&{\tt GTFV} &{\tt GZU} \\
      &&&{\tt l} &{\tt l} &{\tt z} &{\tt y} &&{\tt GTFV} &{\tt GZL} \\
      &&&{\tt vl}&{\tt vl}&{\tt z} &{\tt y} &&{\tt GTFV} &{\tt GZN} \\
      &&&{\tt d} &{\tt u} &{\tt w-}&{\tt y} &&{\tt GTFV} &{\tt GWF} \\
      &&&{\tt u} &{\tt d} &{\tt w+}&{\tt y} &&{\tt GTFV} &{\tt GWF} \\
      &&&{\tt l-}&{\tt vl}&{\tt w-}&{\tt y} &&{\tt GTFV} &{\tt GWF} \\
      &&&{\tt vl}&{\tt l-}&{\tt w+}&{\tt y} &&{\tt GTFV} &{\tt GWF} \\
      &&&{\tt q} &{\tt q} &{\tt g} &{\tt y} &&{\tt GTFV} &{\tt GG}  \\
 VVVT &&&{\tt g} &{\tt g} &{\tt g} &{\tt y} &&{\tt GTV}  &{\tt G}   \\
\hline\hline
\multicolumn{8}{l}{5-point couplings}            &{\tt GT} &{\tt GC} \\
\hline
 VVVVT &&{\tt g}&{\tt g}&{\tt g}&{\tt g}&{\tt y}&&{\tt GTV}&{\tt G}  \\
\hline
\end{tabular}
\caption{List of the coupling constants for each vertex. All the
 particles and the coupling constants are written in the {\tt MG}
 notation. {\tt y} stands for a massive graviton, {\tt f} represents all
 possible fermions, and {\tt v} is the SM gauge bosons ({\tt g},
{\tt a}, {\tt z}, {\tt w}). {\tt GC} is a SM
 coupling constant, while {\tt GT} is a non-renormalizable coupling
 constant defined in each subroutine in Sect.~\ref{sec:helas}.}
\label{couplist}
\end{table}

Finally, we note that since in the ADD model the gravitons are
densely populated and we should sum over their contributions by
modifying the phase space integration in {\tt ME}. In the ADD model,
the spectrum of KK graviton modes can be treated as continuous for
$\delta\leq 6$~\cite{ADD}, and the mass density function is given
by~\cite{Feynr1}
\begin{align}
 \rho(m)=
  S_{\delta-1}\frac{{\overline M}_{\rm Pl}^2}{M_s^{2+\delta}}m^{\delta-1}
  \quad \text{with}\
  S_{\delta-1}=\frac{2\pi^{\delta/2}}{\Gamma(\delta/2)},
\end{align}
where $M_s$ is the ADD model effective scale. Thus, we modify the
phase space generating codes in {\tt ME} to add one more random number
for graviton mass generating and implement the above graviton mass
integration.

\section{Sample results}
\label{sec:sample}

In this section, we present some sample numerical results for the
graviton plus mono-jet and di-jet productions at the LHC, using the new
{\tt HELAS} subroutines and the modified {\tt MG/ME}. 

We use the following jet definition criteria
\begin{align}
 P_T^j>20~{\rm GeV},\ \ |\eta_{j}|< 5,\ \
 R_{jj}= \sqrt{\Delta \eta^2+\Delta \phi^2}> 0.6,
\end{align}
where $\eta$ is the pseudorapidity of the jets and $\phi$ is the
azimuthal angle around the beam direction, and further require
\begin{align}
\eta_{j_1}\cdot\eta_{j_2}<0,\ \ |\eta_{j_1}-\eta_{j_2}|>4.2.
\end{align}
The CTEQ6L1 parton distribution
functions~\cite{Pumplin:2002vw} are employed with the factorization scale
chosen as $\mu_f = {\rm min}(P_T^j)$ of the jets which satisfy the
above cuts. For the QCD coupling, we set it as the geometric mean
value, $\alpha_s=\sqrt{\alpha_s(P_T^{j_1})\cdot\alpha_s(P_T^{j_2}})$.

\begin{figure}
\centering \epsfig{file=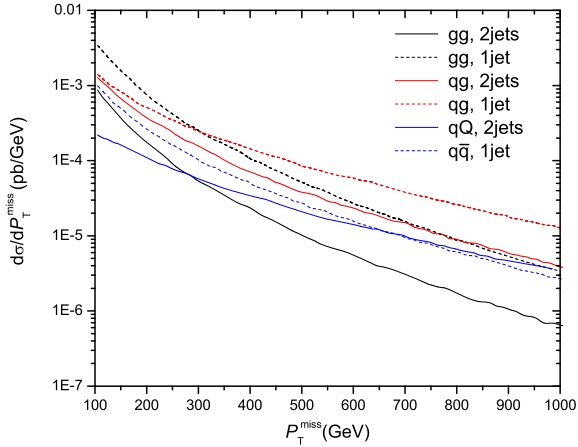,width=1.\columnwidth,clip}
 \caption{\label{fig:add} $P^{{\rm miss}}_T$ dependence of the total
 cross sections for the graviton productions with 1-jet (via the $gg$,
 $qg$ and $q\bar{q}$ channels) and 2-jets (via the $gg$, $qg$ and $qQ$
 channels) in the ADD model at the LHC, where $q,Q=u,d,s,c$.}
\end{figure}

Fig.~\ref{fig:add} shows the $P^{{\rm miss}}_T$ distributions for the
graviton productions with the mono-jet (via the $gg$, $qg$ and
$q\bar{q}$ channels) and with the di-jet (via the $gg$, $qg$ and $qQ$
channels) in the ADD model at the LHC, with $\Lambda=5$~TeV and
$\delta=4$. Note that the unitarity criterion $M_{T_n}<\Lambda$ is
used. See more details in Ref.~\cite{Hagiwara:2008iv}. 

\begin{figure}
\centering \epsfig{file=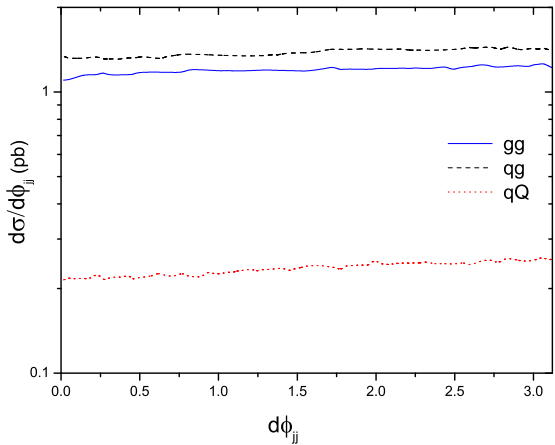,width=1.\columnwidth,clip}
 \caption{\label{fig:rs} Distributions of the azimuthal angle
 separation between the two jets for the first KK graviton plus di-jet
 productions in the RS model at the LHC, via the $gg$, $qg$ and $qQ$
 channels ($q,Q=u,d,s,c$). }
\end{figure}

Fig.~\ref{fig:rs} shows the $\phi_{jj}$ distributions for the first KK
graviton productions with two jets in the RS model at the LHC, via the
$gg$, $qg$ and $qQ$ channels. Here we set $\Lambda=4$~TeV and
$M_{T_1}=1$~TeV. The total cross sections are 0.74, 3.74 and 4.33 pb for
the $qQ$, $gg$ and $qg$ channels, respectively.

\section{Summary}
\label{sec:summary}

In this paper, we have added new {\tt HELAS} subroutines to calculate
helicity amplitudes with massive spin-2 particles (massive
gravitons) to the {\tt HELAS} library. They are coded in such a way that
arbitrary scattering amplitudes with one graviton production and its
decays can be generated automatically by {\tt MG} and {\tt ME},
after slight modifications. All the codes have been tested carefully by
making use of the invariance of the helicity amplitudes under the
gauge and general coordinate transformations.

\begin{acknowledgement}{\textit{Acknowledgement}.}
The authors wish to thank Michel Herquet for teaching about the User
 Model in {\tt MG/ME}.  
Q.L.\ and K.M.\ would like to thank the KEK theory group for the warm
 hospitality, and also the IPMU (Institute for Physics and Mathematics
 of the Universe) for organizing an LHC focus week meeting in December
 2007 where we enjoyed stimulating discussions.  
This work is supported in part by the Core University Program of JSPS,
 the Grant-in-Aid for Scientific Research (No. 17540281) of MEXT, Japan,
 and Deutsche Forschungsgemeinschaft under SFB/TR-9
 ``Computergest\"utzte Theoretische Teilchenphysik''.
\end{acknowledgement}


\end{document}